\documentstyle[11pt,newpasp,twoside,epsf]{article}
\def\edcomment#1{\iffalse\marginpar{\raggedright\sl#1\/}\else\relax\fi}
\reversemarginpar

\begin{document}
\title{Multiwavelength observations of the GRB000615 field}
\author{E. Maiorano$^1$, N. Masetti$^2$, E. Palazzi$^2$, L. 
Nicastro$^3$, E. Pian$^4$, L. Amati$^2$, F. Frontera$^{2,5}$, 
J. in 't Zand$^6$ \& K.Z. Stanek$^7$} 

\affil{$^1$Dept. of Astronomy, Univ. of Bologna;
$^2$IASF/CNR, Bologna;
$^3$IASF/CNR, Palermo;
$^4$INAF - Astron. Obs. of Trieste;
$^5$Dept. of Physics, Univ. of Ferrara;
$^6$SRON, Utrecht;
$^7$Harvard-Smithsonian Center for Astrophysics, Cambridge}

\begin{abstract} 
We report on prompt and afterglow observations of GRB 000615 detected by
\emph{Beppo}SAX. The study of the high-energy prompt event is presented
along with the search for its X--ray and optical afterglow. Spectral fits
seem to suggest a temporal evolution of the GRB prompt emission. We
possibly find evidence for intrinsic $N_{\rm H}$ (at 90\% confidence
level) and for a transient spectral emission feature around 8 keV (at 98\%
confidence level). The X--ray to $\gamma$--ray fluence ratio of 1.73 $\pm$
0.22 is one of the largest among \emph{Beppo}SAX GRBs. A weak X--ray
source is also detected in the MECS, between 1.6 and 4.5 keV, and its
position is compatible with the WFC error box. The behaviour of this
source may be compatible with that of an afterglow. Low significance
signal is detected in the 0.2--1.5 keV at a position consistent with the
WFC and MECS error boxes. The S/N ratio is insufficient to speculate on
the nature of this source. There is no evidence of an optical transient
down to $R \sim 22$.
\end{abstract}

\vspace{-1.0cm}
\section{Introduction}

GRB000615 was simultaneously detected, at 06:17:45 UT of 15 June 2000, by
the \emph{Beppo}SAX GRBM and WFC, with a localization uncertainty of
2$^{\prime}$ (error circle radius), at coordinates (J2000) RA = $15^{\rm
h} 32^{\rm m} 36\fs9$, DEC = $+73^{\circ}49^{\prime}07^{\prime\prime}$
(Gandolfi et al. 2000). \emph{Beppo}SAX NFI observations started $\simeq$
10 hours after the trigger time and lasted 1.44 days. Preliminary NFI data
analysis by Nicastro et al. (2001) showed the presence of two sources, one
in the LECS between 0.1 and 4 keV, and the other in the MECS between 1.6
and 4 keV. It is still not clear whether the MECS and the LECS sources are
really distinct and which (if any) is related to GRB000615. Optical, IR
and radio follow-up observations did not detect any transient in the WFC
error box (Stanek et al. 2000; Pian et al. 2000;  Di Paola et al. 2000;
Frail et al.  2000). Here we present a refined analysis of the high energy
\emph{Beppo}SAX data along with that of optical data of the WFC error box. 

\section{Prompt event}

Figure 1, left panel, shows the WFC (2--10 keV and 10--28 keV) and GRBM
(40--700 keV) light curves of GRB000615. The $\gamma$--ray emission lasted
about 13 s, while the X--ray emission was detected over a substantially
longer interval of 120 s. Moreover, the bulk of the X--ray emission
started $\sim$ 40 s after the $\gamma$--ray peak. In order to perform
time-resolved spectral analysis of the GRB emission in the 2--700 keV
range, we divided the light curves into three time intervals (a, b and c;
see Fig. 1, left panel) of duration 30 s, 30 s and 60 s, respectively. The
spectra integrated over these time bins were analysed separately. In our
fits the $N_{\rm H}$ was fixed at the Galactic value (2.7$\times$10$^{20}$
cm$^{-2}$) unless otherwise stated. The best-fit parameters for the three
spectra are reported in Table~1. 

\begin{figure}
\plotfiddle{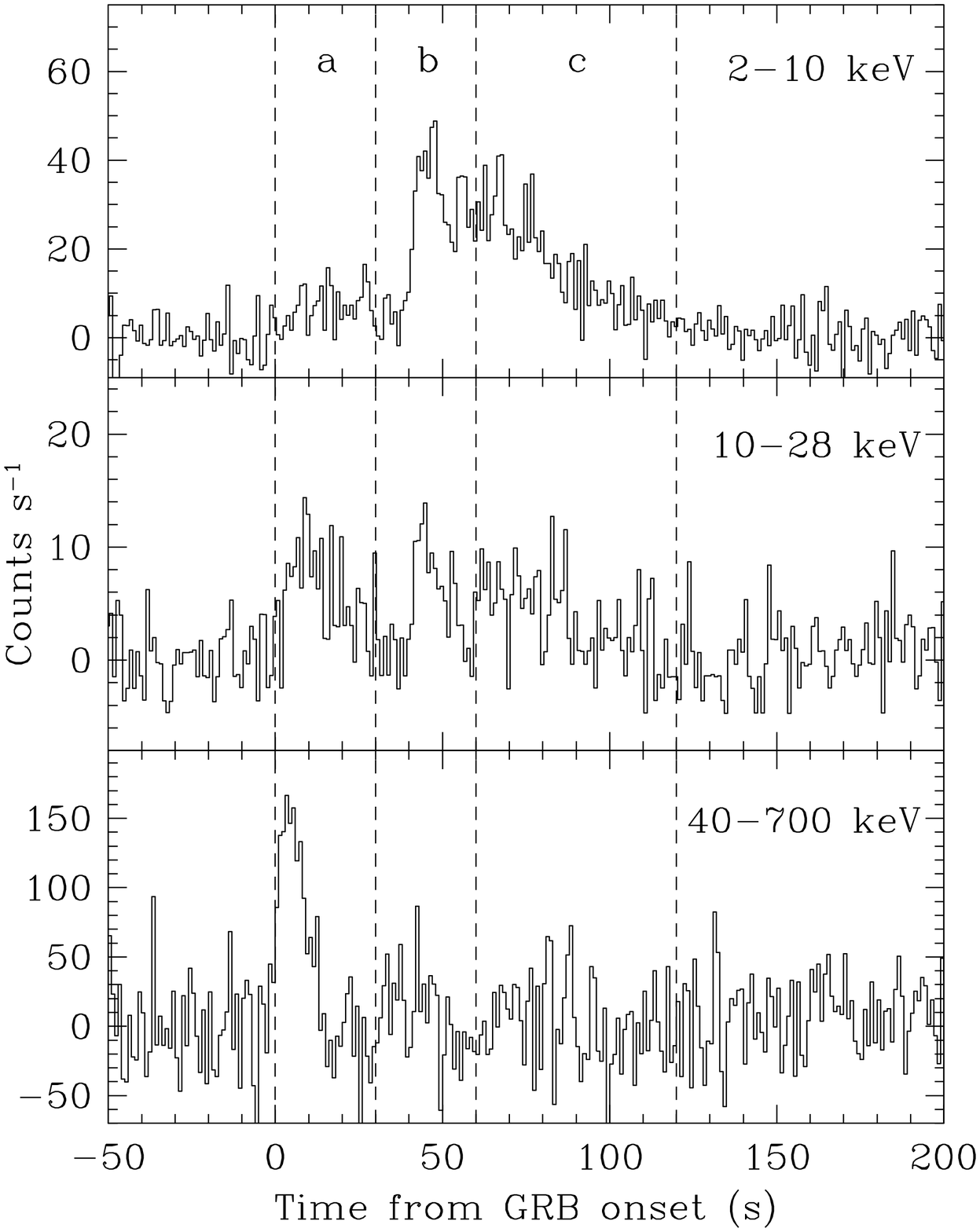}{8cm}{0}{32}{32}{-210}{-5}
\plotfiddle{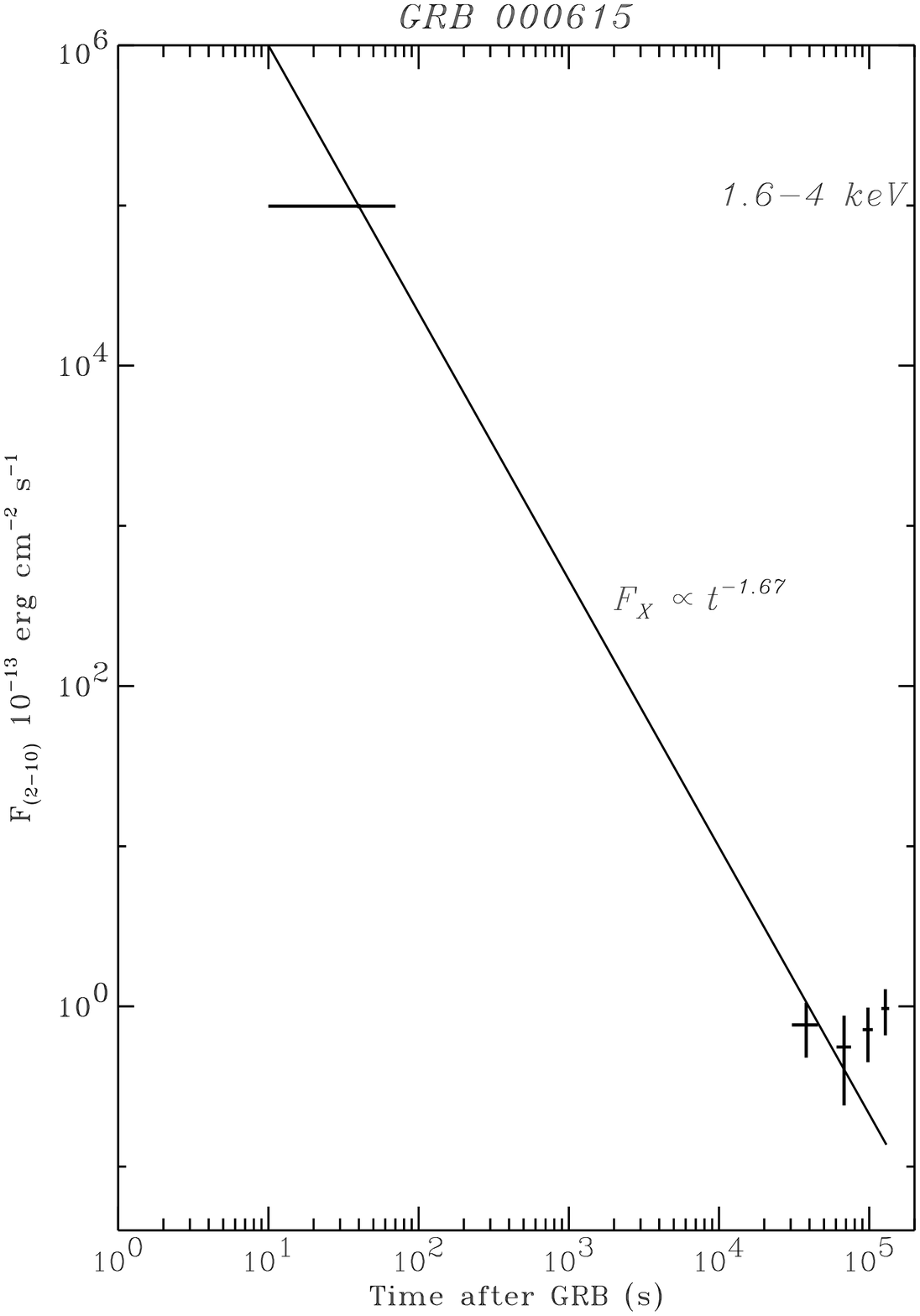}{4cm}{0}{33}{33}{10}{125}
\vspace{-5cm}
\caption{\emph{Left}: WFC (upper and central panels) and GRBM (lower
panel) background subtracted light curves of GRB000615; the temporal
slices a, b and c are marked. \emph{Right}: 2-4 keV flux measured by the
\emph{Beppo}SAX WFC and MECS. A power-law of index 1.7 connnecting the two
measurement is also shown (see text).} 
\end{figure}  

In the first 30 s (slice a) the spectrum was first fitted with a single
power-law (PL) with photon index 1.5 $\pm$ 0.2. However, a 90\%
probability improvement was obtained using a single PL with free $N_{\rm
H}$. In this case, we had a weak suggestion for a large $N_{\rm H}$ value
($\sim$2$\times$10$^{23}$ cm$^{-2}$), as shown in Table~1. We also fitted
these data with the Band model (Band et al. 1993) or with a broken PL, but
the spectral parameters were largely unconstrained.

The WFC data of the second temporal slice (b) were well fitted with a
single PL with index 1.9 $\pm$ 0.3; however, this model were inconsistent
with the GRBM upper limits. When these were accounted for the fit, the PL
index was steeper than 2.12, which was still within 1$\sigma$ from the WFC
data best-fit.
 
The spectrum of slice c was best fitted with a PL of index 2.2 $\pm$ 0.3. 
However, an excess around 8 keV appeared in the residuals. When we added a
narrow Gaussian line to the model, the fit improved at the 98\% confidence
level, but using instead a recombination edge model, the parameters were
poorly constrained. Nevertheless, assuming that this excess was due to the
Fe recombination edge at 9.27 keV, a redshift $z \sim$ 0.13 was derived. 

The 2--10 keV to 40--700 keV fluence ratio of this GRB is 1.73 $\pm$ 0.22,
one of the highest to date found for a \emph{Beppo}SAX GRB (marginally
higher than that of GRB990704, which was 1.52 $\pm$ 0.15; Feroci et al.
2001). 

\vspace{-0.4cm}
\section{The search for the afterglow}

\subsection{X--rays}
Confirming the preliminary results of Nicastro et al. (2001), the analysis
of the MECS data revealed the presence of an uncatalogued steady X-ray
source in the 1.6--4.5 keV interval. Its position was consistent with the
WFC error box of the GRB. The spectrum of this object was fitted with
either a PL with index $\sim$2.3 or with a blackbody having $kT \sim$ 0.6
keV. The average flux in this band was (8.0 $\pm$ 1.6) $\times$ 10$^{-14}$
erg cm$^{-2}$ s$^{-1}$. 

In the LECS image we detected some signal at the center of the WFC error
box with a significance of 2.5 $\sigma$ between 0.2 and 1.5 keV during the
first 60 ks (corresponding to 15 ks of actual on-source time) of the
observation. Its position was 2$'$ away from the center of the MECS source
error box, and therefore marginally consistent with it, considering the
low S/N.  The non-detection of substantial emission in the LECS at the
position of the MECS source suggested that additional absorption ($N_{\rm
H} \sim$ 4$\times$10$^{21}$ cm$^{-2}$) could be present if the PL model
was chosen.  No additional $N_{\rm H}$ was required if, instead, a
blackbody model was assumed.

\begin{table*}[t!]
\caption{Best-fit spectral parameters of the GRB000615 high-energy data.
Errors are at 90\% confidence level. $N_{\rm H}$ in brackets are fixed at 
Galactic value.}
{\small  
\begin{center}
\begin{tabular}{c c c c c c}
 &  &  &  &  & \\
\hline  
\hline 
\noalign{\medskip}
Slice & Model & $N_{\rm H}$ & $\Gamma$ & E$_l$ & $\chi^2$/dof \\
   &  & (10$^{22}$ cm$^{-2}$) &  & (keV) & \\
\noalign{\medskip}
\hline
\hline
A & PL & [0.027] & $1.5\pm{0.2}$ & --- & 5.7/4 \\
  & PL+$N_{\rm H}$ & $21^{+66}_{-19}$ & $1.9^{+0.5}_{-0.3}$ & --- & 2.0/3 \\
  &  &  &  &  &  \\
B & PL & [0.027] & $1.9\pm{0.3}$ & --- & 2.9/8 \\
  &  &  &  &  &  \\
C & PL & [0.027] & $2.2^{+0.4}_{-0.3}$ & --- & 7.9/8 \\
  & PL+GAUSS & [0.027] & $2.6^{+0.5}_{-0.4}$ & $8.2^{+0.8}_{-0.9}$ & 
2.1/6 \\ 
\hline
\hline
\end{tabular} 
\end{center}}
\end{table*}

Figure 1, right panel, shows the 2--4 keV flux of the GRB and of the MECS
source. The plotted line corresponds to a power-law decay with index
$\alpha \simeq$ 1.7 which is typical for X--ray afterglows. The steadiness
of the flux from the source detected in the MECS is intriguing and could
suggest that this was not the GRB afterglow; nevertheless this unusual
behaviour occurred at least in another case, GRB970508 (Piro et al. 1998),
in which a rebursting laid on a powerlaw decay was observed. 

\subsection{Optical} We performed an accurate analysis of $R$-band data
(Stanek et al. 2000) acquired at FLWO on June 15 and 16, 2000; the results
confirmed that 4 hours after the GRB trigger no optical afterglow was
present in the WFC error box down to $R \sim$ 22, i.e. slightly deeper
than the previous limit ($R \sim$ 21.5) determined from these data. 

\vspace{-0.3cm}
\section{Summary}

GRB000615 is probably the X--ray richest GRB observed by {\it BeppoSAX},
with a fluence ratio $S_X/S_{\gamma} = 1.73 \pm 0.22$. We found marginal
evidence of spectral evolution in the prompt event: during the first 30 s,
the spectrum shows very marginal evidence of absorption in addition to the
Galactic one, which may be due to a high density circumburst medium.
Between 60 and 120 s after the trigger, an emission feature is present at
98\% confidence level, which can be interpreted as a Fe recombination edge
at $z \sim$ 0.13.

An uncatalogued and non-significantly variable source is detected with the
MECS in the 1.6--4.5 keV energy range. The 2--4 keV fluxes measured by the
WFC during the prompt event and by the MECS during the NFI pointing can be
connected assuming a power-law decay of index 1.7, which is typical for
X-ray afterglows. While the steadiness of the MECS source flux is uncommon
in afterglows, we note, as remarked before, that this behavior is
reminiscent of the short-term variability exhibited by the X--ray
afterglow of GRB970508.

A transient 2.5$\sigma$ signal is present below 1.5 keV in the LECS image,
at a position marginally consistent with the MECS source, but fully
consistent with WFC error box. Although this latter circumstance may
suggest its association with the GRB, the low significance of the
detection prevents us to establish its reality and its relationship with
the afterglow.

No optical afterglow down to $R\sim22$ is present and this might suggest
that GRB000615 is a ``dark" burst which likely occurred in a high
density environment.


\end{document}